\documentclass[reprint, preprintnumbers, superscriptaddress, amsfonts, aps, prl, nofootinbib]{revtex4-2}

\usepackage{amsmath}
\usepackage{mathtools}
\usepackage{wasysym}

\def\l{\left}
\def\r{\right}

\def\ch{\mathrm{ch}}
\def\dN{{dN_\ch/d\eta}}
\def\dNcr{\l(dN_\ch/d\eta\r)^{1/3}}

\begin{document}
\count\footins = 1000
\title{The Multiplicity Dependence of Pion Interferometry in Hydrodynamics}

\author{Christopher Plumberg}
\email{christopher.plumberg@gmail.com}
\thanks{ORCID: https://orcid.org/0000-0001-6678-3966}
\affiliation{Illinois Center for Advanced Studies of the Universe, Department of Physics, University of Illinois at Urbana-Champaign, Urbana, IL 61801, USA}


\begin{abstract}
Understanding the origins of collective, fluid-like behavior in ultrarelativistic nuclear collisions constitutes one of the biggest open challenges in the field.  In this Letter, it is argued that certain features in the multiplicity dependence of the source sizes extracted using pion interferometry may be understood quite naturally if small systems evolve hydrodynamically at sufficiently large multiplicities.  Interferometry may therefore provide a baseline for probing and constraining the nature of collective behavior in nuclear collisions.
\end{abstract}

\date{\today}

\maketitle


\noindent {\sl \underline{1. Introduction.}} It is by now widely recognized that ultrarelativistic nuclear collisions exhibit collective, fluid-like behavior (or `collectivity') \cite{Nagle:2018nvi, Adolfsson:2020dhm}. The current challenge is to understand what produces this behavior.  To date, many different explanations of collectivity have been proposed \cite{Martinez:2018tuf, Wertepny:2020jun, Lin:2015ucn, Bierlich:2017vhg, Sjostrand:2018xcd, Kurkela:2018ygx, Schenke:2010nt, *Gale:2012rq, *Gale:2013da, Shen:2014vra, Weller:2017tsr, Schenke:2019pmk}.  In order to experimentally discriminate between these proposals, one needs a set of observables which can probe the consequences of collective behavior in highly specific and non-trivial ways.

Collectivity is commonly defined by strong space-momentum correlations \cite{Ollitrault:1992bk, *Ollitrault:1995dy}.  As such, it should manifest itself in both momentum-space and coordinate-space observables \cite{Voloshin:2003ud}.  An ideal way to probe the latter in nuclear collisions is provided by Hanbury Brown--Twiss (HBT) interferometry and the observables derived from it, the `HBT radii' \cite{Wiedemann:1999qn, Lisa:2005dd}.

The HBT radii reflect collectivity in a number of ways, including their dependence on the transverse pair momentum $K_T$ and the system's charged multiplicity $\dN$.  However, although a fairly clean and extensive theoretical structure exists for understanding the former \cite{Akkelin:1995gh, Heinz:1999rw}, somewhat less attention has been paid to quantitatively extracting insights from the latter, although important work has been done on this as well \cite{Lisa:2005dd, Kisiel:2008ws}.

For a thermalized, hydrodynamic medium, one expects the spatial volume of the system to scale approximately linearly with the multiplicity $\dN$, suggesting that the individual radii scale linearly with $\dNcr$ \cite{Lisa:2005dd}.  This rough expectation has been abundantly confirmed in experimental data \cite{Aamodt:2011kd, Adam:2015pya, Adam:2015vna}.  However, there are two significant features of the $\dNcr$-dependence in the radii which initially appear to contradict what one expects on the basis of hydrodynamic models.  First, one observes that the slope of the $\dNcr$-scaling in each radius varies between large and small collision systems.  I refer to this variation as a \textit{slope non-universality} in the scaling of the radii across different size systems.  Second, one finds that different radii exhibit different slopes in the \textit{same} collision system.  In the widely used out-side-long ($o$, $s$, $l$) coordinate system \cite{Podgoretsky:1982xu}, for instance, with primary radii $R_o$, $R_s$, and $R_l$, the $R_o$ slope tends to be considerably smaller than the $R_s$, $R_l$ slopes in small systems, whereas all three are roughly comparable in magnitude in larger systems \cite{Kisiel:2011jg, Graczykowski:2014hoa, Plumberg:2020jod}.  There is thus a \textit{slope hierarchy} between the respective radii whose magnitude depends on the collision system in which it is measured.  Taken together, these two features -- the hierarchy and non-universality in the slopes of the radii with $\dNcr$ -- appear initially to stand in tension with the hydrodynamic expectation that all radii should scale in similar ways across collision systems.

The goal of this Letter is to show that these experimental trends are in fact automatic consequences of hydrodynamic behavior and arise naturally in systems which exhibit anomalously strong collective flow.  A simple model is used to explore heuristically the qualitative behavior of the different HBT radii in hydrodynamics as a function of system size and multiplicity and to show what this implies about the underlying space-time evolution of the systems in question.  A systematic and quantitative analysis of the HBT radii and their multiplicity dependence may therefore elucidate the nature of collectivity in crucial ways and open up novel avenues for exploring the mechanisms which drive the dynamical evolution of ultrarelativistic nuclear collisions.

\begin{figure*}
\centering
\includegraphics[width=\linewidth, keepaspectratio]{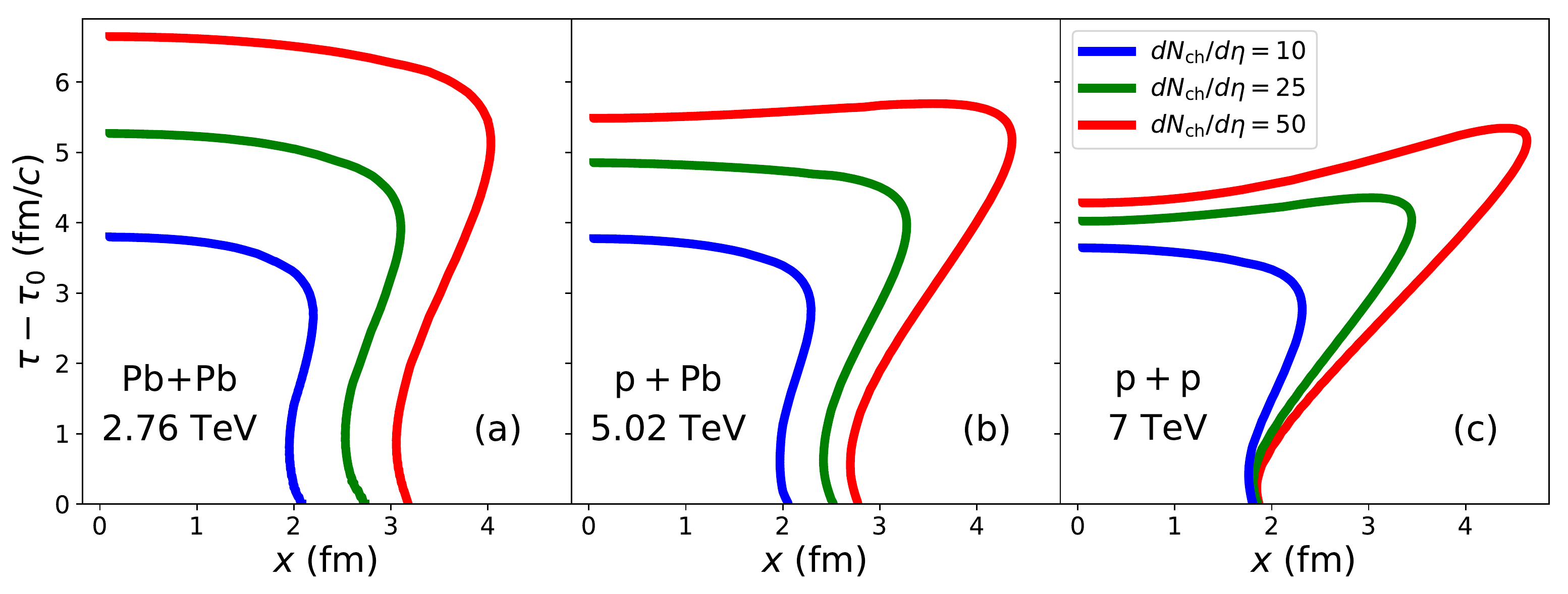}
\caption{Comparison of freeze-out surface slices (at $y=0$) in Pb+Pb (a), p+Pb (b), and p+p (c) collision systems, for different fixed multiplicities of $dN_\ch/d\eta=10$ (blue, innermost contours), 25 (green, middle contours), and 50 (red, outermost contours).  As noted in \cite{Heinz:2019dbd}, the comoving volume of the freeze-out surface is fixed by the multiplicity, but the space-time volume enclosed by the freeze-out surface reflects the highly system-dependent spatial geometry and strength of collective flow. \label{Fig:FOsurfaces}}
\end{figure*}

\noindent {\sl \underline{2. Methodology.}} The starting point for HBT analyses is the two-particle correlation function:
\begin{equation}
	C(\vec{p}_1, \vec{p}_2)
	= \frac{E_1 E_2 \frac{dN}{d^3p_1 d^3p_2}}
	       {\l(E_1 \frac{dN}{d^3p_1}\r)
	        \l(E_2 \frac{dN}{d^3p_2}\r)}.
	\label{2pCF_def}
\end{equation}
It is constructed ideally to reduce to unity in the absence of Bose-Einstein or Fermi-Dirac correlations.  In this Letter, this correlation function is computed and studied for pions produced by three different systems as functions of $\dNcr$: p+p collisions at 7 TeV, p+Pb collisions at 5.02 TeV, and Pb+Pb collisions at 2.76 TeV.

In terms of the relative momentum $q$ and the pair momentum $K$, defined by
\begin{equation}
	q = p_1 - p_2,\quad K=\frac{1}{2}\l(p_1 + p_2\r),
\end{equation}
one may parameterize the correlation function \eqref{2pCF_def} in a form which is Gaussian in $\vec{q}$:
\begin{equation}
  C_{fit}(\vec{q}, \vec{K})
   = 1 + \lambda(\vec{K}) \exp\l( - \sum_{i,j \in \l\lbrace o,s,l \r\rbrace} R^2_{ij}(\vec{K}) q_i q_j \r),
   \label{fitCF}
\end{equation}
where $(o,s,l)$ label the respective axes of the out-side-long coordinate system \cite{Podgoretsky:1982xu}, and I have specified to the case of pion (i.e., Bose-Einstein) correlations.  The strength of the Bose-Einstein enhancement is absorbed into the normalization factor $\lambda(\vec{K})$, which may deviate from unity in the presence of resonance decays \cite{Wiedemann:1996ig} and coherent pion production \cite{Sinyukov:1994en, *Sinyukov:2012ut, *Akkelin:2011zz, *Shapoval:2013jca}.  Here, I assume purely chaotic and thermal pion emission for simplicity, so that $\lambda(\vec{K})=1$.  The inverse square widths of the enhancement are parametrized by the HBT radii $R^2_{ij}(\vec{K})$, which quantify the effective sizes (or `homogeneity lengths' \cite{Akkelin:1995gh}) of regions in the system which dominate pair production at a given $\vec{K}$, and thereby convey a mixture of spatial and temporal information regarding the evolution and particle emission process of nuclear collisions.

In the case of a system whose pions are emitted according to a phase-space distribution $S$ (or `source function' \cite{Heinz:1996bs}) which is Gaussian in its space-time dependence,\footnote{This is a reasonable assumption for the thermal pion source used here \cite{Plumberg:2016sig}.} one can show that the following `pocket relations' hold exactly \cite{Heinz:2004qz}:
\begin{eqnarray}
R^2_s &=& \l< \tilde{x}_s^2 \r> \label{R2s_GSA}\\
R^2_o &=& \l< \tilde{x}_o^2 \r>
          - 2\beta_T\l< \tilde{x}_o \tilde{t} \r>
          + \beta_T^2\l< \tilde{t}^2 \r> \label{R2o_GSA}\\
R^2_l &=& \l< \tilde{x}_l^2 \r>
          - 2\beta_L\l< \tilde{x}_l \tilde{t} \r>
          + \beta_L^2\l< \tilde{t}^2 \r>, \label{R2l_GSA}
\end{eqnarray}
where all averages are taken with respect to $S$ and the shifted coordinates $\tilde{x}^\mu$ and pair velocity $\vec{\beta}$ are given by \cite{Heinz:1999rw}
\begin{equation}
\tilde{x}^\mu = x^\mu - \l< x^\mu \r>,\quad \vec{\beta} = \frac{\vec{K}}{K^0}
	\approx \frac{\vec{K}}{\sqrt{m_\pi^2 + \vec{K}^2}}. \label{beta_def}
\end{equation}
Each of the terms in \eqref{R2s_GSA}-\eqref{R2l_GSA} characterizes a particular spatiotemporal dimension of the particle source $S$ and is thus termed a ``source variance" \cite{Plumberg:2015eia}.   Although the radii reported here are extracted by fitting \eqref{fitCF} to the correlation function \eqref{2pCF_def} as computed within a hydrodynamic approach, the relations \eqref{R2s_GSA}-\eqref{R2l_GSA} establish a connection between the $R^2_{ij}$ and the source variances which quantify the space-time structure of $S$, thereby reflecting the `freeze-out hypersurface' at which hydrodynamics is terminated and the system is converted to particles \cite{Heinz:2004qz}.

For this work, the boost-invariant hydrodynamic framework iEBE-VISHNU \cite{Shen:2014vra} was used to model p+p collisions at 7 TeV, p+Pb collisions at 5.02 TeV, and Pb+Pb collisions at 2.76 TeV, using smooth, event-averaged MC-Glauber initial conditions \cite{Loizides:2014vua, *Bozek:2019wyr}.  The correlation function \eqref{2pCF_def} was evaluated numerically in terms of integrals over each system's freeze-out hypersurface with respect to the corresponding source function $S$.  It was then fit as a function of $\vec{q}$ and $\vec{K}$ to the parameterization \eqref{fitCF}, thereby yielding the radii $R^2_{ij}(\vec{K})$.  Further details of the hydrodynamic implementation used and the subsequent extraction of the radii have been given in Ref.~\cite{Plumberg:2020jod}.


\begin{figure}
\centering
\includegraphics[width=\linewidth]{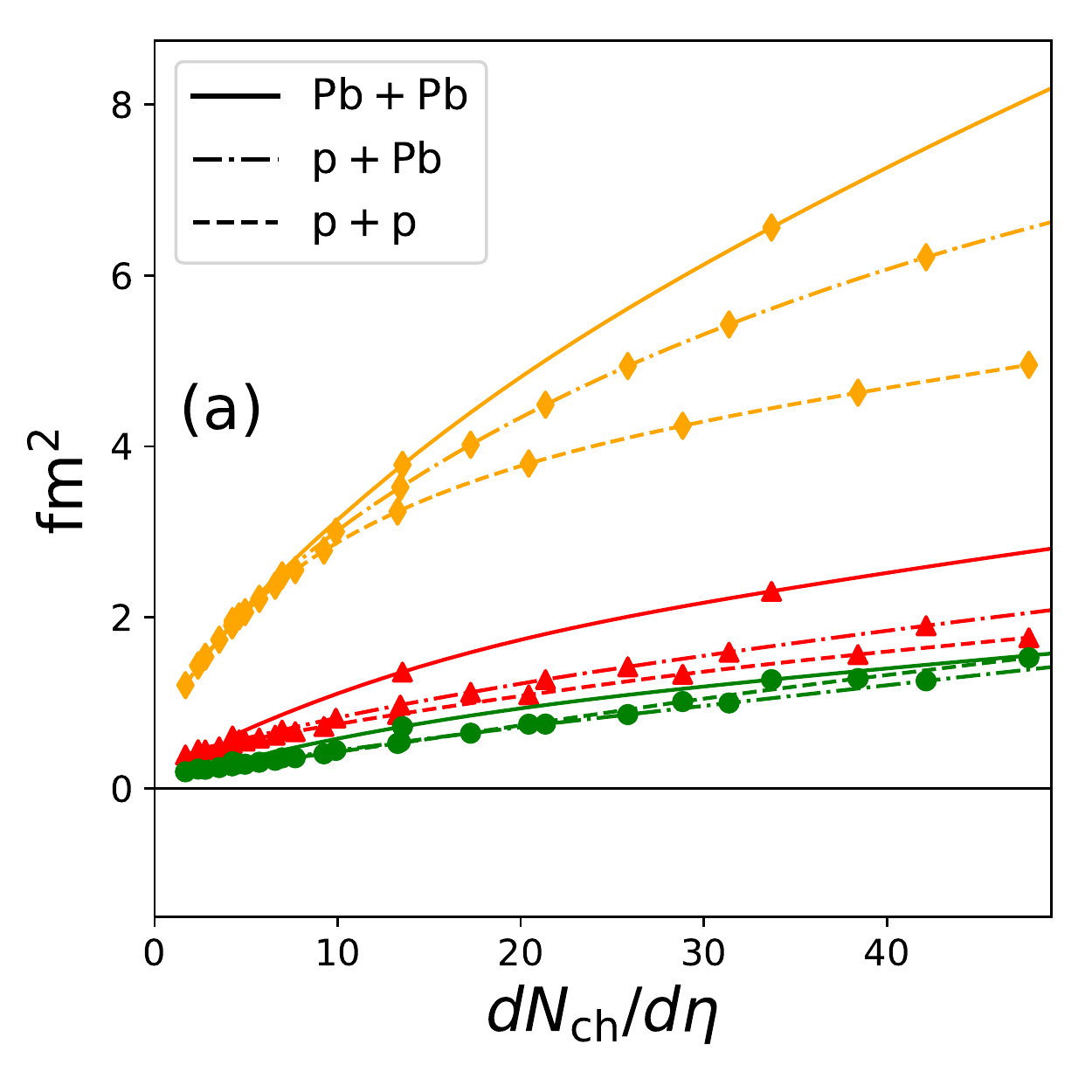}
\includegraphics[width=\linewidth]{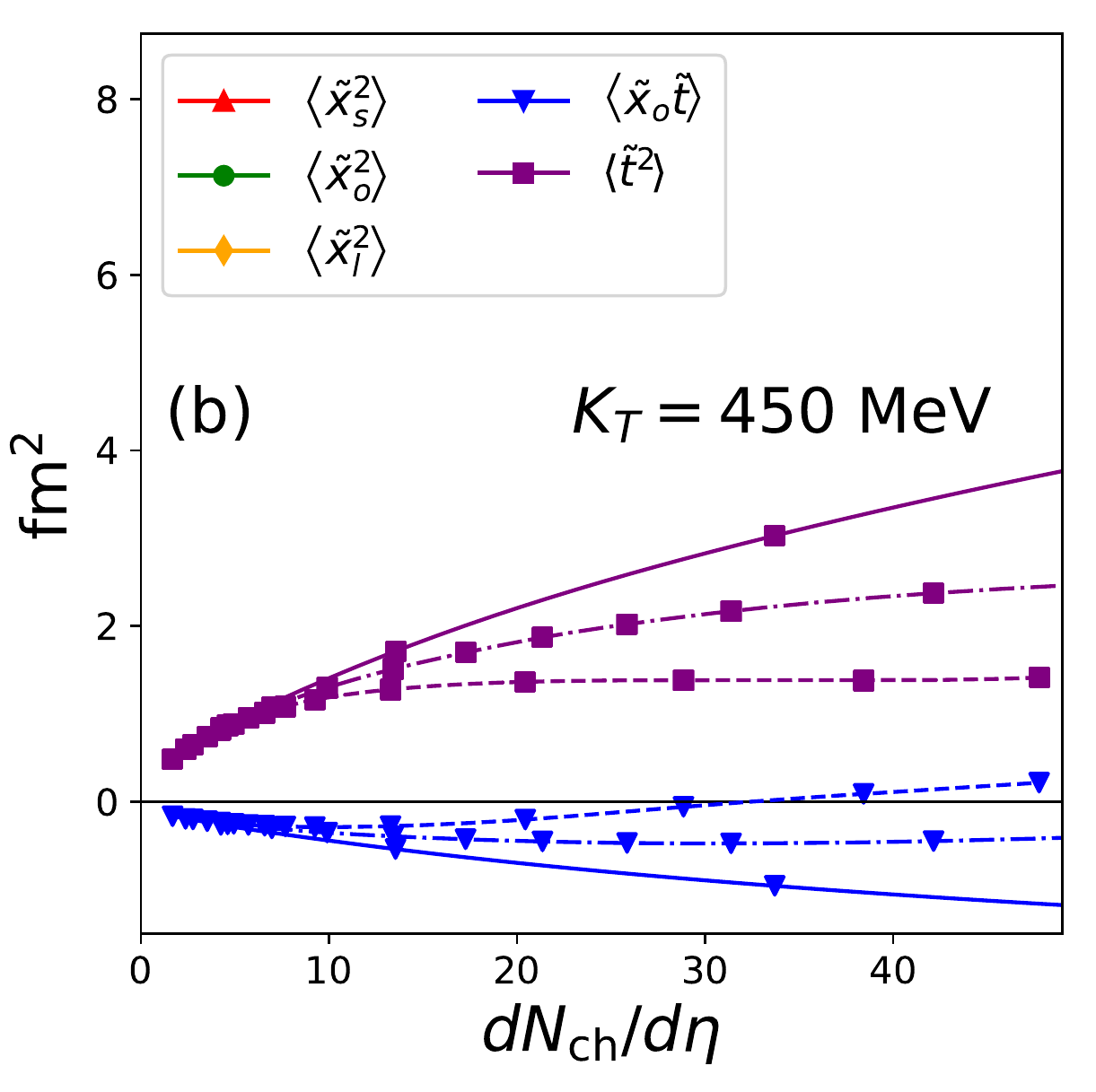}
\caption{The evolution of the space-time geometry, as reflected in the various terms entering the pocket relations \eqref{R2s_GSA}-\eqref{R2l_GSA} with $\dN$.  The geometric terms ($\l< \tilde{x}_s^2 \r>$, $\l< \tilde{x}_o^2 \r>$, $\l< \tilde{x}_l^2 \r>$) scale monotonically with $\dN$ in both large and small systems, whereas the terms sensitive to the temporal structure ($\l< \tilde{t}^2 \r>$, $\l< \tilde{x}_o \tilde{t} \r>$) exhibit radical differences in $\dN$-scaling between different collision systems. \label{Fig:SVscaling}}
\end{figure}

\begin{figure*}
\includegraphics[width=\linewidth, keepaspectratio]{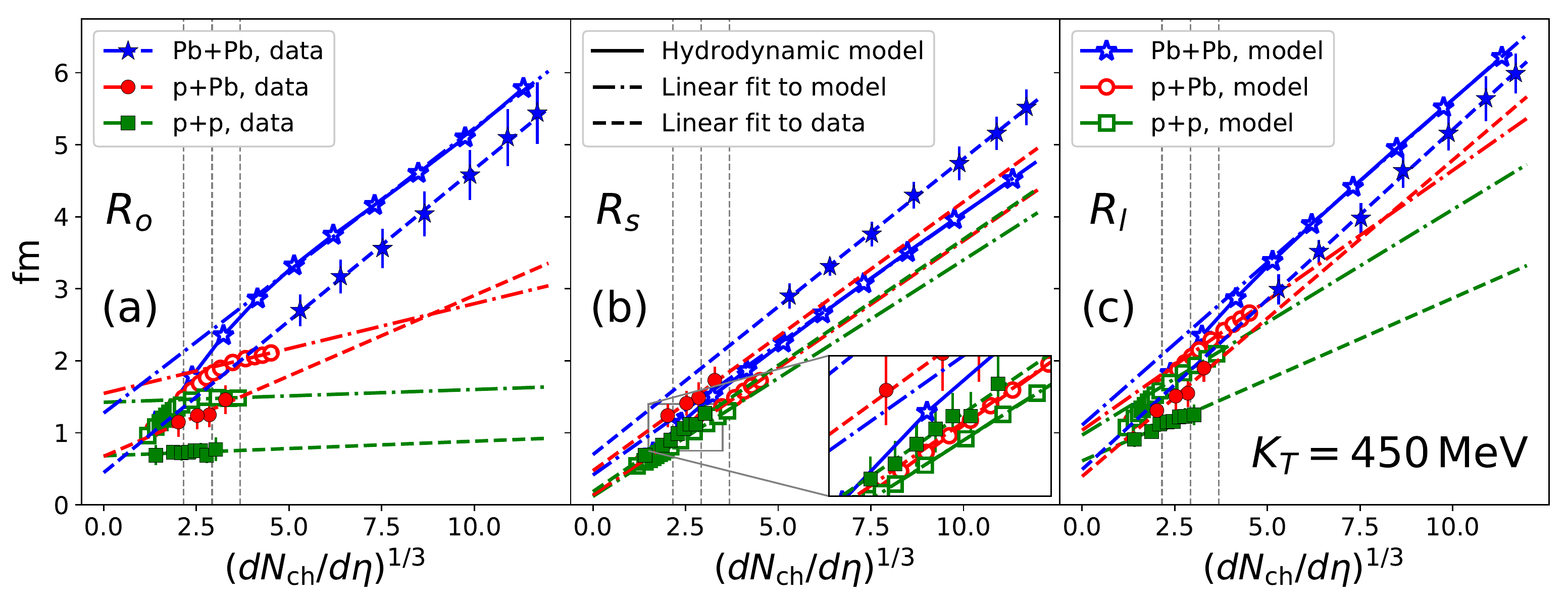}
\caption{Hydrodynamic calculations of the $\dNcr$-dependence in $R_o$ (a), $R_s$ (b), and $R_l$ (c), compared with corresponding experimental measurements in p+p at 7 TeV \cite{Aamodt:2011kd}, p+Pb at 5.02 TeV \cite{Adam:2015pya}, and Pb+Pb at 2.76 TeV \cite{Adam:2015vna}.  Dashed lines and the corresponding symbols correspond to model calculations; dash-dotted lines and their symbols apply similarly to the experimental results.  The three vertical dashed lines correspond to the three multiplicities displayed in Fig.~\ref{Fig:FOsurfaces}. \label{Fig:model_to_data_comparison}}
\end{figure*}

\noindent {\sl \underline{3. Results.}}
I now consider, using the hydrodynamic model just described, how the effects of collective flow may be manifested in the multiplicity dependence of pion interferometry.  I first show the freeze-out hypersurfaces themselves, followed by the source variances, and finally the radii as functions of $\dNcr$.

{\sl 3a. Freeze-out hypersurfaces}. In Fig.~\ref{Fig:FOsurfaces}, I show the freeze-out hypersurfaces obtained from each simulated collision system for different fixed multiplicities $\dN = 10$, 25, 50.  One finds that large systems and small systems differ dramatically in their multiplicity dependence: whereas the former change mainly by a global rescaling of the enclosed space-time volume, with only minor changes to the overall shape, the latter simultaneously grow larger and become distorted in their shape with increasing $\dN$ \cite{Kisiel:2008ws}.  The distortions arising in small systems are driven primarily by enhanced collective flow, a consequence of the fact that small systems are created with substantially larger initial density gradients than large systems \cite{Heinz:2019dbd, Plumberg:2020jod}.

In the highest multiplicity p+p and p+Pb collisions, one further notices a \textit{reversal} of the order in which freeze out occurs in Pb+Pb: the center tends to freeze out \textit{first} in small systems [cf.~Fig.~\ref{Fig:FOsurfaces}(b)-(c)], followed by the edges, whereas in large systems the center freezes out \textit{after} the edges [cf.~Fig.~\ref{Fig:FOsurfaces}(a)].  This reversal in turn generates a strong, \textit{positive} correlation in small systems between a given pion's $x_o$ emission coordinate and the proper time $\tau$ at which the pion was produced.  These features are again consequences of strong transverse expansion and lead to corresponding effects on the source geometry as probed by pion interferometry.

{\sl 3b. Source variances}. The evolution with $\dN$ of the freeze-out hypersurfaces determines the corresponding behavior of the source function $S$ and its space-time structure.  This structure can be probed quantitatively by considering the various terms entering the pocket relations \eqref{R2s_GSA}-\eqref{R2l_GSA}.  In Fig.~\ref{Fig:SVscaling}, we show the five terms which contribute to the radii in the LCMS frame, where $\beta_L \equiv 0$ \cite{Csorgo:1991ej, Chapman:1994ax}.

Fig.~\ref{Fig:SVscaling}(a) reveals a monotonic growth in the geometric terms ($\l< \tilde{x}_s^2 \r>$, $\l< \tilde{x}_o^2 \r>$, $\l< \tilde{x}_l^2 \r>$) with $\dN$.  This growth reflects the corresponding growth with multiplicity in the spatial sizes of the systems shown in Fig.~\ref{Fig:FOsurfaces}; each freeze-out hypersurface must have the same co-moving volume at a fixed multiplicity, irrespective of the collision's enclosed space-time volume \cite{Heinz:2019dbd}.  Fig.~\ref{Fig:SVscaling}(b), by contrast, shows non-monotonic behavior in the temporal structure ($\l< \tilde{t}^2 \r>$, $\l< \tilde{x}_o \tilde{t} \r>$) of the source function $S$.  This, too, is a direct consequence of the multiplicity scaling seen in Fig.~\ref{Fig:FOsurfaces}.  One observes that in small systems at large multiplicities, the emission duration $\l< \tilde{t}^2 \r>$ flattens dramatically while the correlation term $\l< \tilde{x}_o \tilde{t} \r>$ changes sign from negative to positive; both effects result from the elongated `wing-like' structures visible in Fig.~\ref{Fig:FOsurfaces}(b) and (c).  By combining the scaling of the source variances together with the pocket relations \eqref{R2s_GSA}-\eqref{R2l_GSA}, one is thus led to expect on the basis of hydrodynamics that the $R_o$ scaling with $\dNcr$ in small systems will be much weaker than that of $R_s$ or $R_l$, whereas in large systems one expects to see a more uniform scaling in all radii \cite{Plumberg:2020jod}.

{\sl 3c. HBT radii}. The multiplicity scaling of the source function $S$ and its space-time structure ultimately determine the HBT radii extracted the correlation function \eqref{2pCF_def}.  This is shown in Fig.~\ref{Fig:model_to_data_comparison} for the three radii and collision systems under consideration, for both the model hydrodynamic calculations and the corresponding experimental measurements, as functions of $\dNcr$.  For reference, vertical gray lines (dashed) have been added to indicate the three $\dN$ values shown in Fig.~\ref{Fig:FOsurfaces}.  To guide the eye, fit lines have been added to each experimental (dashed) and hydrodynamic (dash-dotted) dataset.  In the latter case, the fits are performed only in the five highest multiplicity bins where the effects of the wing-like freeze-out structure in small systems become apparent.

One finds in these results strong confirmation of the intuition provided by the above analysis of the pocket relations \eqref{R2s_GSA}-\eqref{R2l_GSA}.  There is an obvious non-universality in the slopes of $R_o$ in large and small systems: for $K_T = 450$ MeV, the slope magnitudes in Fig.~\ref{Fig:model_to_data_comparison}(a) are strongly ordered by collision size, and the hydrodynamic trends are in quite good qualitative agreement with those in the data.  By constrast, virtually no splitting is visible in the $R_s$ slopes [Fig.~\ref{Fig:model_to_data_comparison}(b)], and the experimental slopes are again well reproduced by the model.  Some larger model-to-data discrepancies are seen in the slopes of $R_l$ [Fig.~\ref{Fig:model_to_data_comparison}(c)] which are likely due to several simplifying assumptions in the model used here \cite{Plumberg:2020jod}.\footnote{The model generally overestimates the magnitudes of $R_o$ and $R_s$ but underestimates $R_s$, which is a consequence of the fact that preequilibrium flow has not been included here \cite{Pratt:2008qv, Plumberg:2020jod}.}  Most importantly, the qualitative agreement between the model and the data is remarkably good, given the simplicity of the model.


\noindent {\sl \underline{4. Discussion.}} Hydrodynamics naturally predicts enhanced collective flow in small systems due to larger initial density gradients over those present in large systems.  This leads to measurable effects in the $\dNcr$-scaling of the HBT radii obtained from pion interferometry.  Most notably, one observes the development of both a slope hierarchy between different radii in the same system and a slope non-universality when comparing the same radius across different collision systems.  Both features are evident in the experimental data, which are qualitatively similar to the model results presented here.

The effects are strongest in $R_o$, which exhibits a strong splitting across different systems at intermediate $K_T$, while effects on $R_l$ are qualitatively similar but less pronounced.  Essentially no splitting is seen in the slopes of $R_s$, which is relatively insensitive to the temporal structure of the source distribution $S$ \cite{Chapman:1994ax}.

Hydrodynamics thus provides a powerful and natural explanation of the disparate slopes in the various radii over a range of collision systems, by connecting it with the radical differences in transverse flow as a function of system size.  The simple model used here (which is described more fully in Ref.~\cite{Plumberg:2020jod}) provides quantitatively lacking, but qualitatively reasonable, agreement with experimental data.  Because $R_s$ and $R_l$ as measured in the LCMS frame are dominated by geometric aspects of the source function $S$, they scale in a roughly monotonic fashion with $\dNcr$.  On the other hand, because $R_o$ generally contains a mixture of both spatial and temporal information regarding $S$, the effects of violent collective expansion in small systems lead to strong distortions of the hydrodynamic freeze-out structure and much weaker resultant scaling with $\dNcr$.

The $\dNcr$-dependence of the HBT radii thus presents a highly sensitive test for models of collectivity in nuclear collisions.  When combined with suitably complementary momentum-space observables, it may even offer the ability to discriminate between existing models of collectivity.  Future studies will explore these findings in the context of more realistic hydrodynamic models.

\noindent {\sl Acknowledgments.}
The author gratefully acknowledges the use of computing resources from Computing resources from both the Minnesota Supercomputing Institute (MSI) at the University of Minnesota and the Ohio Supercomputer Center \cite{OhioSupercomputerCenter1987}.  C.~P.~ acknowledges support from the CLASH project (KAW 2017-0036) and from the US-DOE Nuclear Science Grant No. DE-SC0019175.

\end{document}